\documentclass{iau}

\usepackage{graphicx}

\usepackage[square,numbers]{natbib}

\newcommand{\apj}{ApJ}

\newcommand{\aap}{A\&A}

\newcommand{\mnras}{MNRAS}

\title[] 
{The evolution of DARWIN: current status\\ of wind models for AGB stars}

\author[]   
{Sara Bladh$^1$}

\affiliation{$^1$Division of Astronomy and Space Physics, Department of Physics and Astronomy, Uppsala University, Box 516, 751 20 Uppsala, Sweden
\\email: sara.bladh@physics.uu.se}

\pubyear{2018}
\volume{343}  
\jname{Why Galaxies Care About AGB Stars: A Continuing Challenge through Cosmic Time}
\editors{F. Kerschbaum, M. Groenewegen \& H. Olofsson, eds.}
\begin{document}

\maketitle

\begin{abstract}
The slow, dense winds observed in evolved asymptotic giant branch (AGB) stars are usually attributed to a combination of dust formation in the dynamical inner atmosphere and momentum transfer from stellar photons interacting with the newly formed dust particles. Wind models calculated with the DARWIN code, using this mass-loss scenario, have successfully produced outflows with dynamical and photometric properties compatible with observations, for both C-type and M-type AGB stars. Presented here is an overview of the DARWIN models currently available and what output these models produce, as well as future plans.

\end{abstract}

\firstsection 

\begin{figure}
\centering
\includegraphics[width=0.45\textwidth]{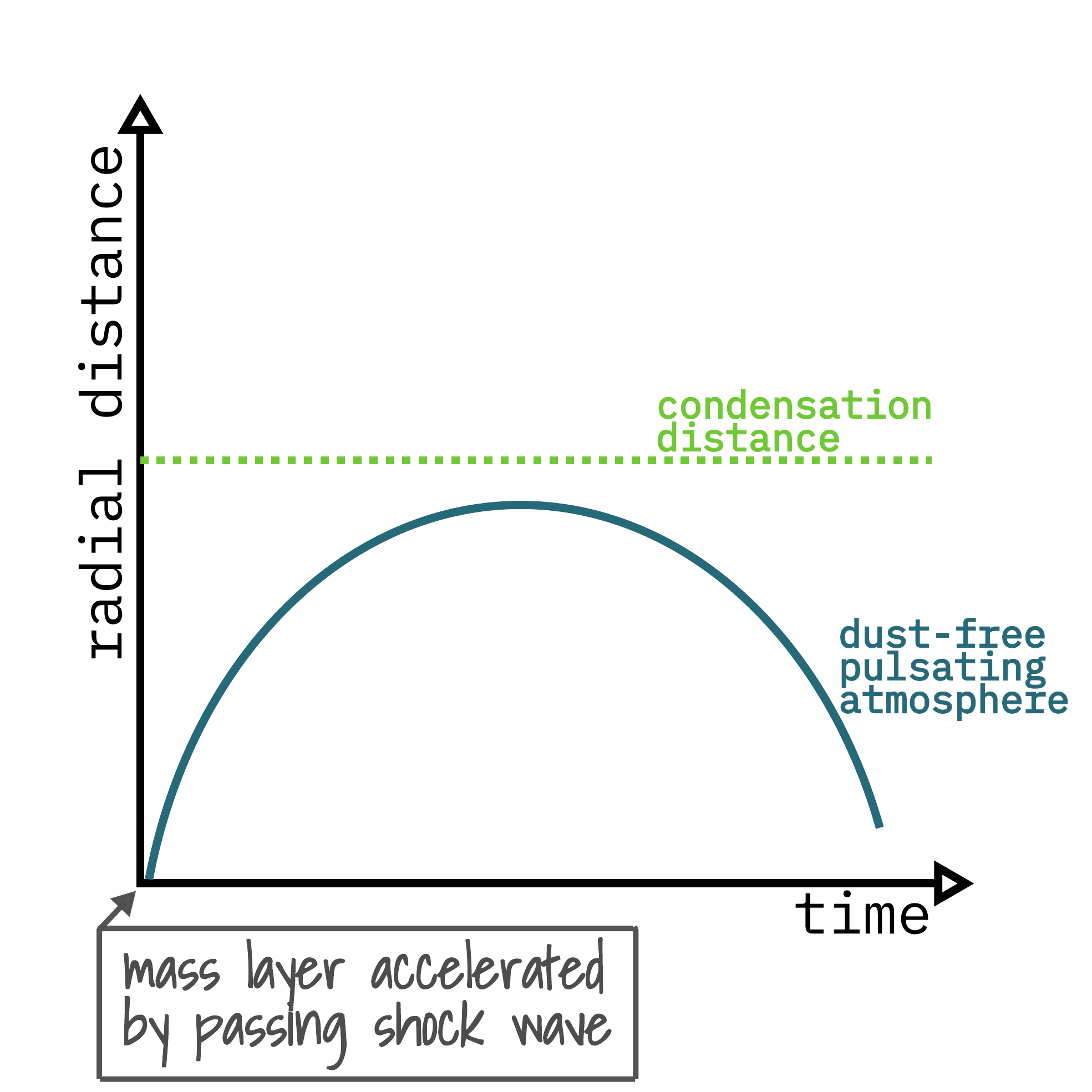}
\includegraphics[width=0.45\textwidth]{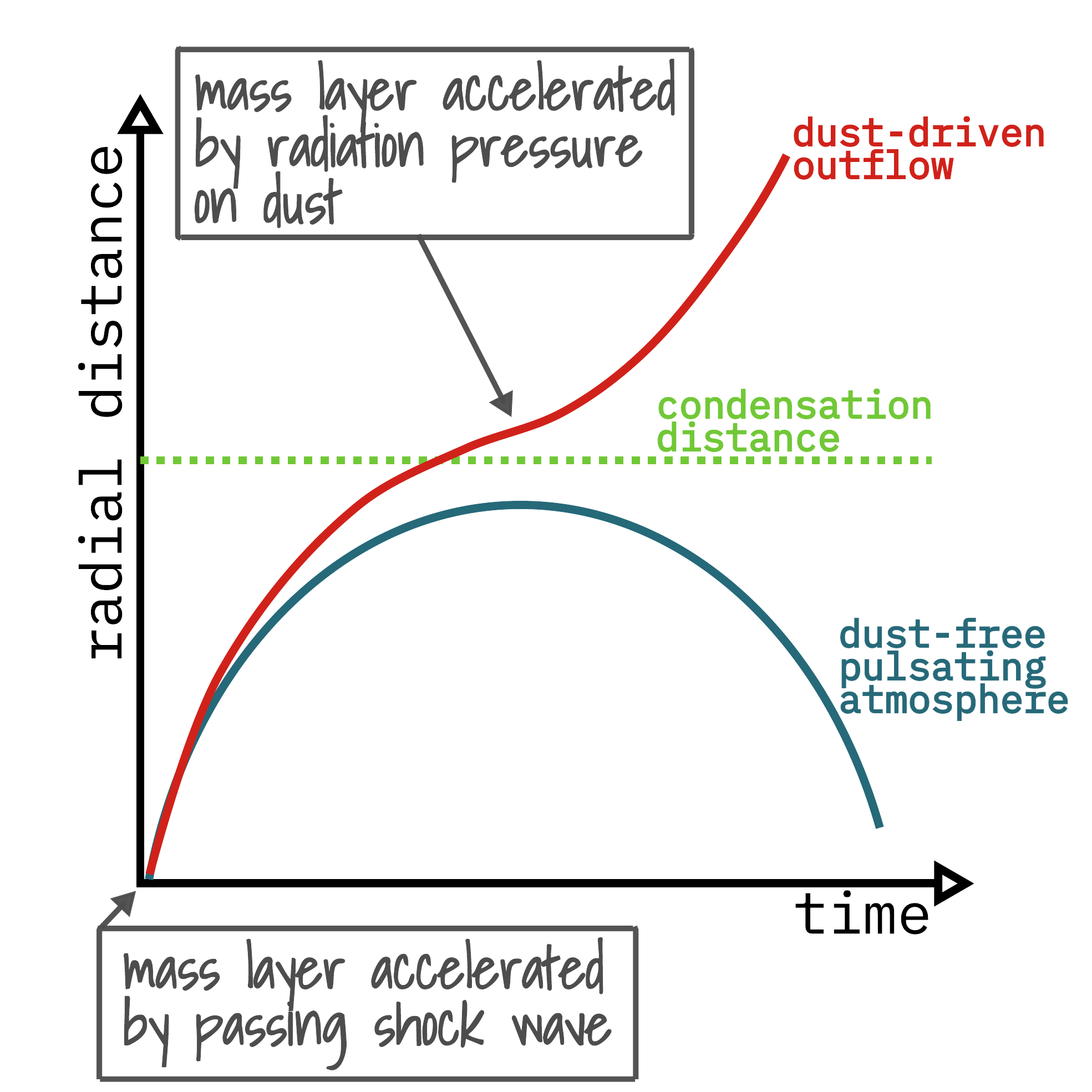}
   \caption{Visualisation of the two-stage process of the pulsation-enhanced dust-driven outflow (PEDDRO) scenario. Image by courtesy of S. Liljegren.}
    \label{f_twostage}
\end{figure}

\section{Introduction}
Stellar winds in evolved AGB stars are generally considered to be pulsation-enhanced dust-driven outflows \citep[PEDDRO, see][for a more detailed explanation]{hoefner18}. This mass-loss scenario is built on a two-stage process, where in the first stage, the contracting and expanding photosphere of the star triggers shock waves that propagate through the steep density gradient of the atmosphere. The shock waves transfer kinetic energy, originating in the pulsations, into the atmosphere, leading to almost ballistic trajectories of the gas. The result is an extended atmosphere with layers of enhanced density at higher altitudes, and consequently, cooler temperatures: an environment favourable to dust formation. In the second stage of this wind scenario, momentum is transferred to the newly formed dust particles by absorption and scattering of the numerous stellar photons reaching the dust formation zone. Friction between the accelerated dust particles and the surrounding gas then triggers a general outflow. A visualisation of the two-stage process of pulsation-enhanced dust-driven outflows is given in Fig.~\ref{f_twostage}.

\section{The DARWIN code}
\begin{figure}
\centering
\includegraphics[width=\textwidth]{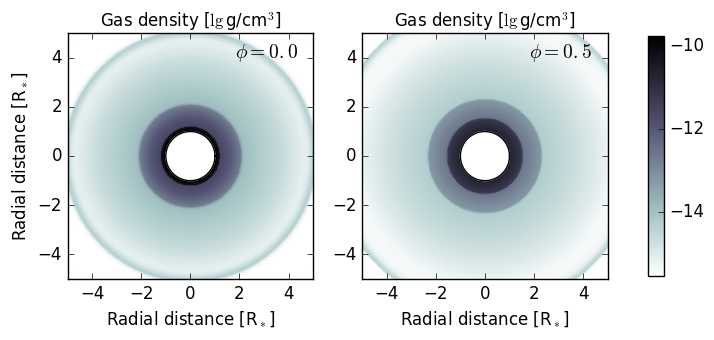}\\
\includegraphics[width=\textwidth]{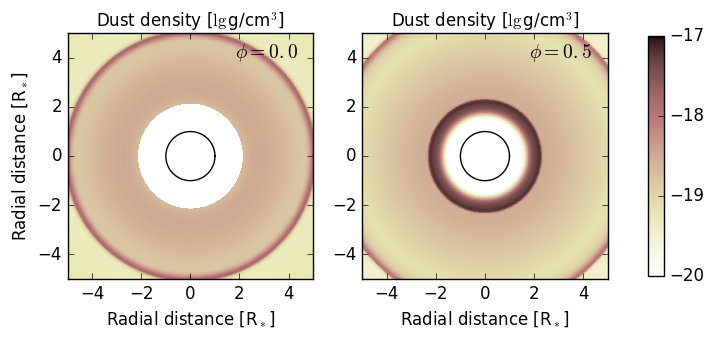}\\
\includegraphics[width=\textwidth]{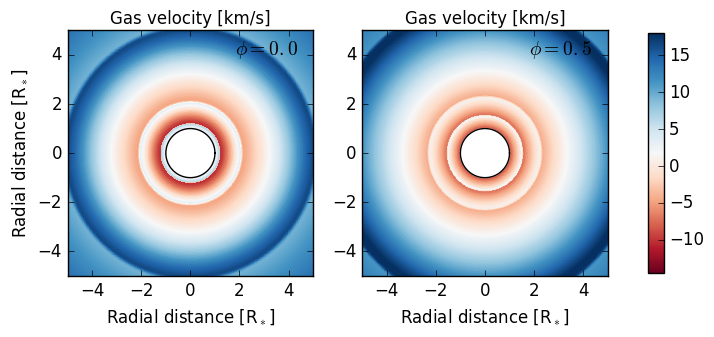}
   \caption{Cross-sectional snapshots of a DARWIN model for a C-type AGB star, with $M_*=1\,\mathrm{M}_{\odot}$, $\log L_*=4.00\,\mathrm{L}_{\odot}$, $T_*=2600\,$K, $u_{\mathrm{p}}=2$\,km/s, and $[\mathrm{Fe/H}]=-1.0$. The upper, middle and lower panels show gas density, dust density, and velocity of the gas, respectively. The left and right panels show cross-sectional snapshots at maximum ($\phi=0.0$) and minimum ($\phi=0.5$) luminosity, respectively. The solid black line indicates the radius of the star, calculated from Stefan-Boltzmann's law.}
    \label{f_structure}
\end{figure}

The atmospheres and winds of AGB stars are modeled using the 1D radiation-hydro\-dynamic code DARWIN (Dynamic Atmosphere and Radiation-driven Wind models based on Implicit Numerics). The DARWIN code produces time-dependent radial structures of the atmospheres and winds of AGB stars by simultaneously solving the hydrodynamic equations (conservation of mass, momentum, and energy), frequency-dependent radiation transfer and time-dependent grain growth. These wind models are  spherically symmetric, with an inner boundary situated just below the photosphere and an outer boundary at about 25 stellar radii in models that develop a wind. The initial structure is a hydrostatic model atmosphere, characterised by the fundamental stellar parameters (current stellar mass $M_\star$, stellar luminosity $L_\star$, effective temperature $T_\star$) and chemical composition. The variability of the star is simulated by a sinusoidal variation (described by the pulsation period $P$, velocity amplitude $u_{\mathrm{p}}$, and a scaling factor $f_{\mathrm{L}}$) of radius and luminosity at the inner boundary. The amplitude of this variation is gradually ramped up during the first pulsation cycles, thereby turning the initial hydrostatic atmosphere into a dynamical atmosphere. Figure~\ref{f_structure} shows atmospheric structures of a DARWIN model at maximum and minimum luminosity (at $\phi$\,=\,0.0 and 0.5, respectively). The different panels show gas density (upper panels), dust density (middle panels) and velocity (lower panels). Note the gas moving inwards and outwards in the inner atmosphere, before the material is accelerated outwards by radiation pressure on dust. 

A detailed description of the DARWIN code can be found in \cite{hoefner2016}. The output from these wind models compares well with observed wind properties and photometry \citep{nowotny10,nowotny11,eriksson14, bladh2013, bladh2015}, illustrating their capability to reproduce the overall momentum transfer and spectral energy distribution of AGB stars. 

\section{The available DARWIN models}
\begin{figure}
\centering
\includegraphics[width=0.49\textwidth]{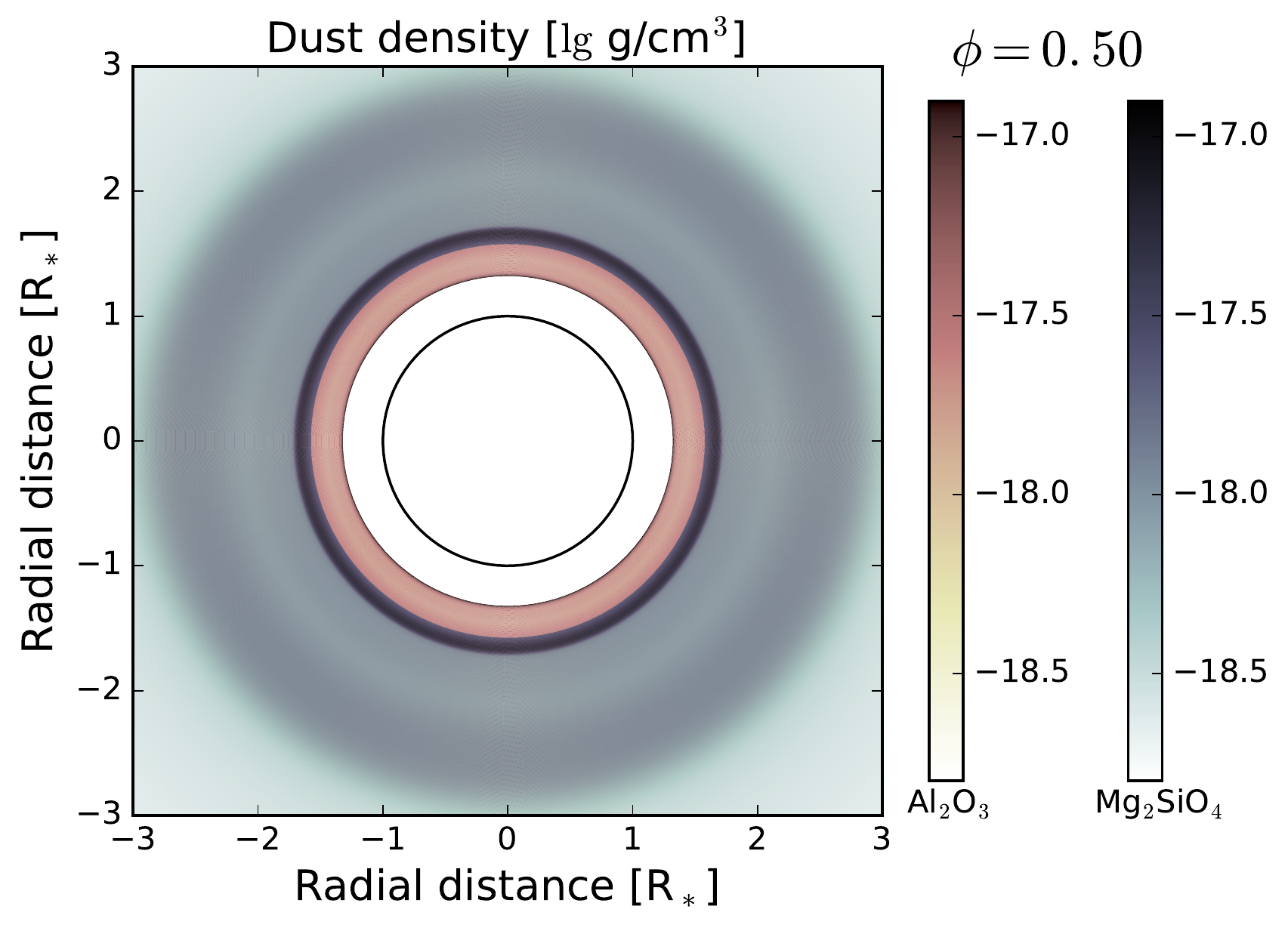}
\includegraphics[width=0.49\textwidth]{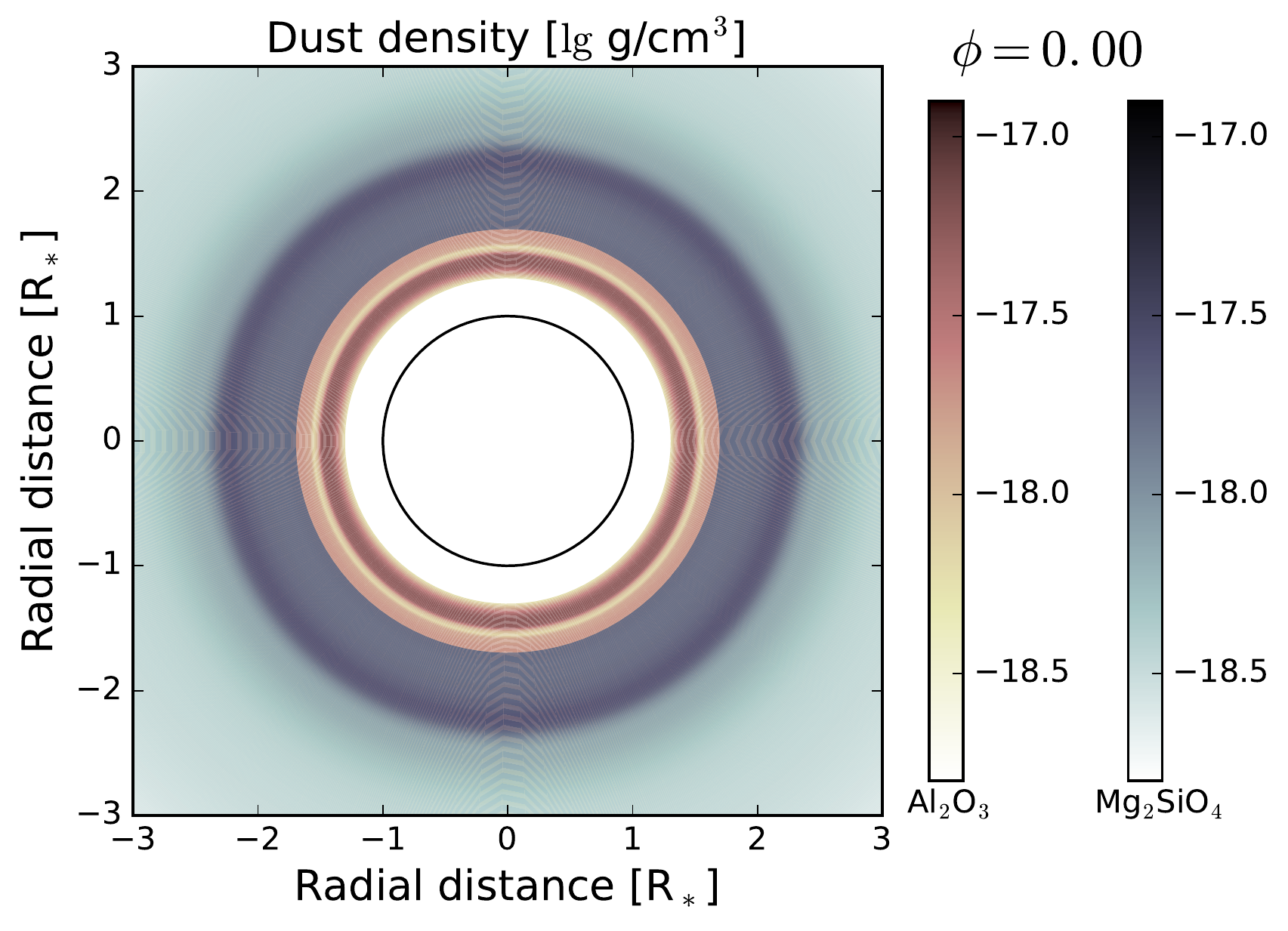}
   \caption{Cross-sectional snapshots at maximum (left panel, $\phi=0.0$) and minimum (right panel, $\phi=0.5$) luminosity, showing the dust density of an oxygen-rich DARWIN model with grain growth of both Al$_2$O$_3$ and Mg$_2$SiO$_4$. The model parameters are $M_*=1\,\mathrm{M}_{\odot}$, $\log L_*=3.70\,\mathrm{L}_{\odot}$, $T_*=2700\,$K, and $u_{\mathrm{p}}=3$\,km/s. The solid black line indicates the radius of the star, calculated from Stefan-Boltzmann's law.}
    \label{f_ostructure}
\end{figure}

The DARWIN code exists in two versions; one for M-type AGB stars, where the wind is driven by photon scattering on Fe-free silicate grains of sizes comparable to the wavelength of the flux maximum, and one for C-type AGB stars, where the wind is driven by photon absorption on amorphous carbon grains. In both versions of the code the dust grains are assumed to be spherical and the optical properties are calculated using Mie theory.

The DARWIN models for C-type AGB stars include a description for the nucleation of amorphous carbon grains, based on classical nucleation theory. The carbon abundance is treated as a free parameter,  as carbon may be dredged-up during the thermal pulses when AGB stars evolve. The relevant quantity for wind-driving in carbon stars is the carbon excess, C-O, since it is the amount of free carbon in the atmosphere, i.e. the carbon that is not bound in CO molecules, that is important for dust formation. There exist two generations of DARWIN models for C-type AGB: the first generation of models assumed that dust particles are small compared to the wavelengths at which the stars emit most of their stellar flux and the dust opacity is calculated in the small particle limit (SPL). In the second generation of models this assumption is relaxed and size-dependent dust opacities (SDO) are used instead.
The first large model grid for C-type AGB stars was published by \cite{mattsson10}, providing mass-loss rate, wind velocities, and dust yields for a wide range of stellar parameters (also see contributed talk ``Calibrating TP-AGB stellar models and chemical yields through resolved stellar populations in the SMC''  by Pastorelli at IAU Symp. 343). This publication also includes a mass-loss routine that can be used in stellar evolution modelling. A slightly smaller grid of DARWIN models for C-type AGB stars, with updated opacities, was published by \cite{eriksson14}. It provides mass-loss rates, wind velocities, and dust yields, but also photometry and a spectral library. Both of these grids assumed small particle limit when calculating the dust opacities. A new grid similar to the one published in \cite{eriksson14}, but with size-dependent opacities, will soon be published (see the poster ``DARWIN C-star model grid with new dust opacities'' by Eriksson at IAU Symp. 343), following up on the pilot study by \citep{mattsson11}

DARWIN models of M-type AGB stars include pre-existing seed particles that start to grow when the thermodynamical conditions are favourable, since the nucleation processes in oxygen-rich environments is still not fully understood \citep[see e.g.][]{gail16,gobrecht16}.  The oxygen-rich models all include size-dependent dust opacities, which is crucial when the wind is driven by photon scattering. The DARWIN models for M-type AGB stars also come in two flavours. The standard model \citep{hoefner2008, bladh2015} includes time-dependent dust growth of Fe-free silicates (Mg$_2$SiO$_4$), but there is also a modified version \citep{hoefner2016} that includes time-dependent dust growth of both Al$_2$O$_3$ and Mg$_2$SiO$_4$ grains. In these modified models Al$_2$O$_3$ forms the core of the grain, with Mg$_2$SiO$_4$ condensing in a mantle surrounding the Al$_2$O$_3$-cores. Figure~\ref{f_ostructure} shows cross-sectional snapshots of the dust density of a DARWIN model with grain growth of both Al$_2$O$_3$ and Mg$_2$SiO$_4$. Note that  Al$_2$O$_3$ condenses in the close vicinity of the star, with Mg$_2$SiO$_4$ growing on top of the Al$_2$O$_3$-core a bit further out. The first grid of DARWIN models for M-type AGB stars was published by \cite{bladh2015}. This grid only included one solar mass models. A new extensive grid of oxygen-rich DARWIN models, including models with different masses, will soon be published \citep{bladh2018a}.

\begin{figure*}
\centering
\includegraphics[width=\textwidth]{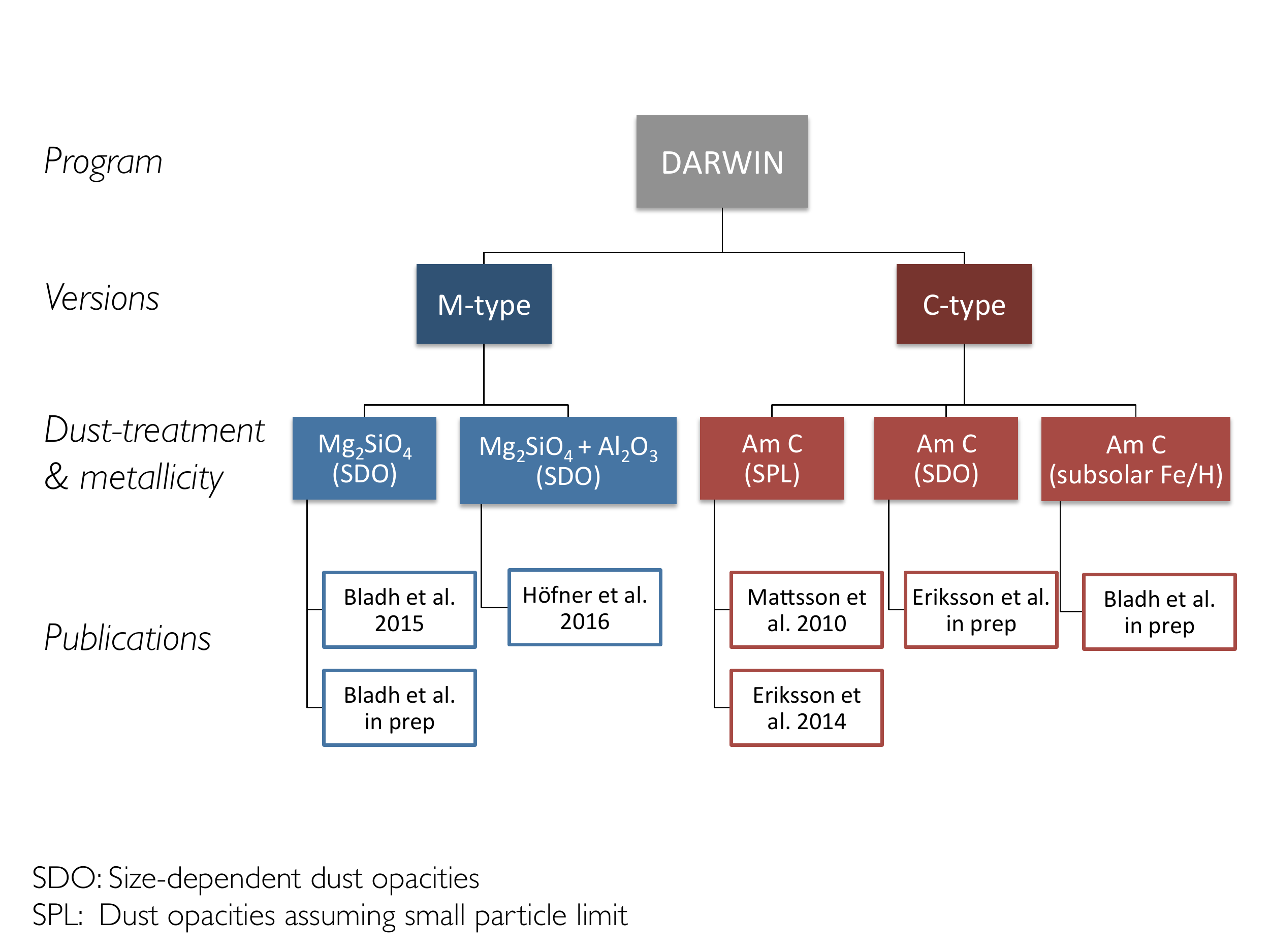} 
\caption{Overview of DARWIN models currently available or soon to be published.}
\label{f_darwin}
\end{figure*}

There is also an ongoing effort to explore wind models in sub-solar metallicity environments. A first exploratory work for carbon-rich DARWIN models was published by \cite{mattsson08}, and a more extensive investigation will be published soon in \cite{bladh2018b}, showing that if carbon excess is used as defining parameter, then the mass-loss rates at sub-solar metallicities agree well with those at solar metallicities for the same stellar parameters. From this it can be concluded that as long as AGB stars manage to dredge up sufficient carbon from the interior, they can contribute to dust production even at lower metallicities. Furthermore, this indicates that stellar evolution models can use the C-type DARWIN solar grids for mass-loss rates at sub-solar metallicities if they select models with the same carbon excess. Figure~\ref{f_darwin} gives an overview of the DARWIN models currently available or soon to be published. 

\section{Output from DARWIN models}
\begin{figure*}
\centering
\includegraphics[width=\textwidth]{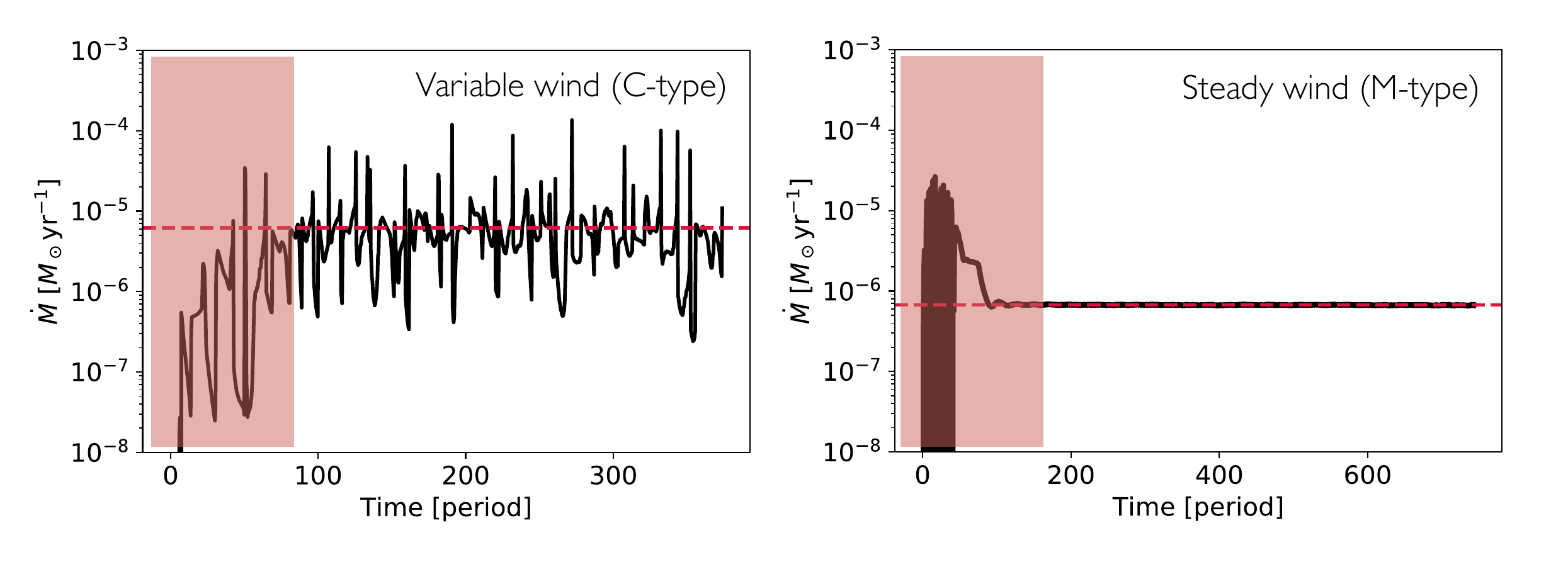} 
\caption{Mass-loss rates at the outermost layers of two DARWIN models as a function of time. The averaged mass-loss rates is marked with a red line. The shaded area indicates the early pulsation periods that are excluded in order to avoid transient effects when calculating the average mass-loss rate.}
\label{f_mlr}
\end{figure*}

The DARWIN models consist of long time-series of snapshots of the atmospheric structures. From these time-series we can derive wind properties, i.e., the wind velocity and mass-loss rate at the outermost layers of the model, averaged over typically hundreds of pulsation periods. The early pulsation periods are excluded to avoid transient effects of ramping up the amplitude. Figure~\ref{f_mlr} shows the wind properties for two such time-series. The left panel is an example of a stellar wind from a DARWIN model where the mass-loss rate over time oscillates quite a lot around the average mass-loss rate (indicated by the red line). The right panel shows a DARWIN model where the mass-loss rate over time instead is extremely steady.

\begin{figure*}
\centering
\includegraphics[width=0.49\textwidth]{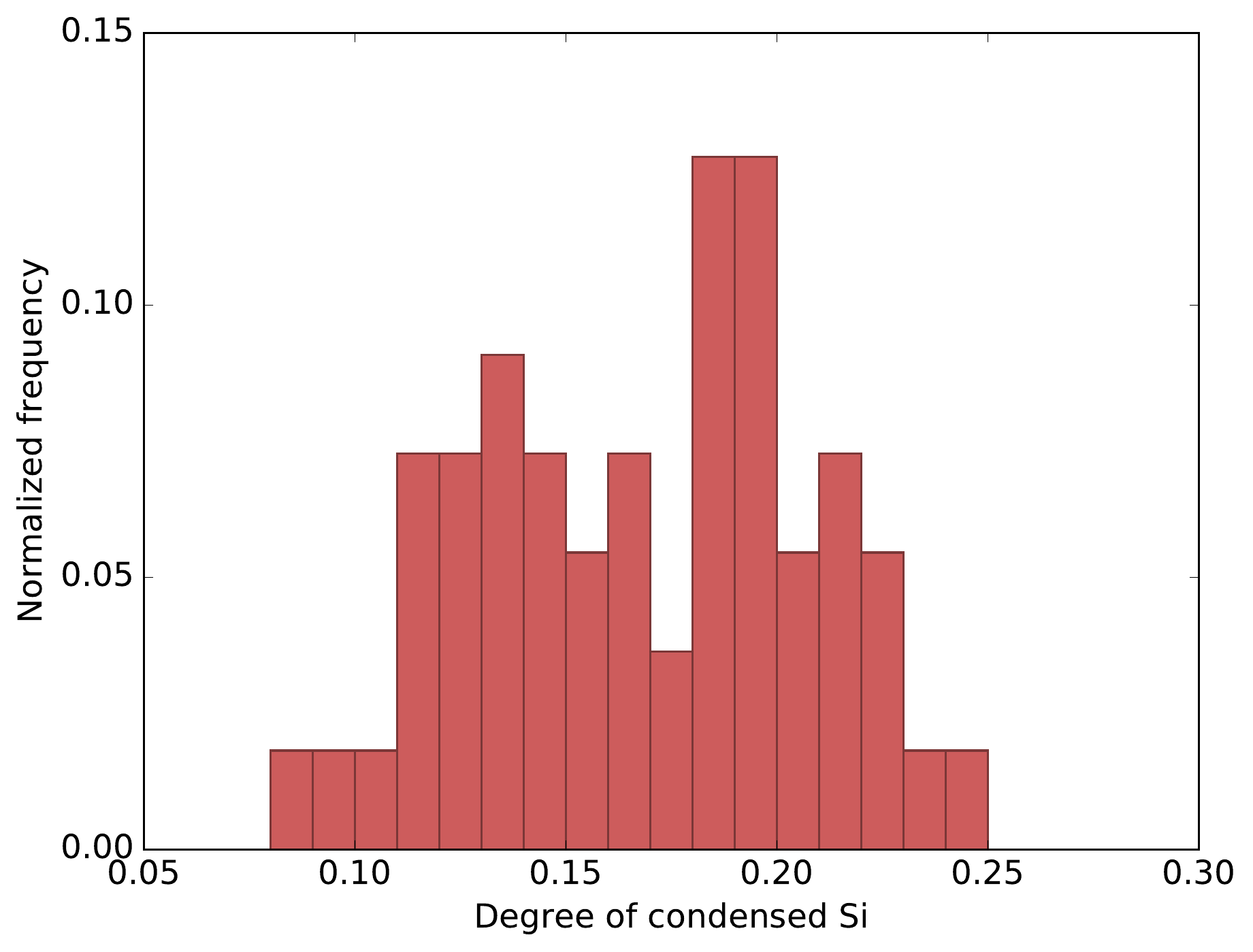} 
\includegraphics[width=0.49\textwidth]{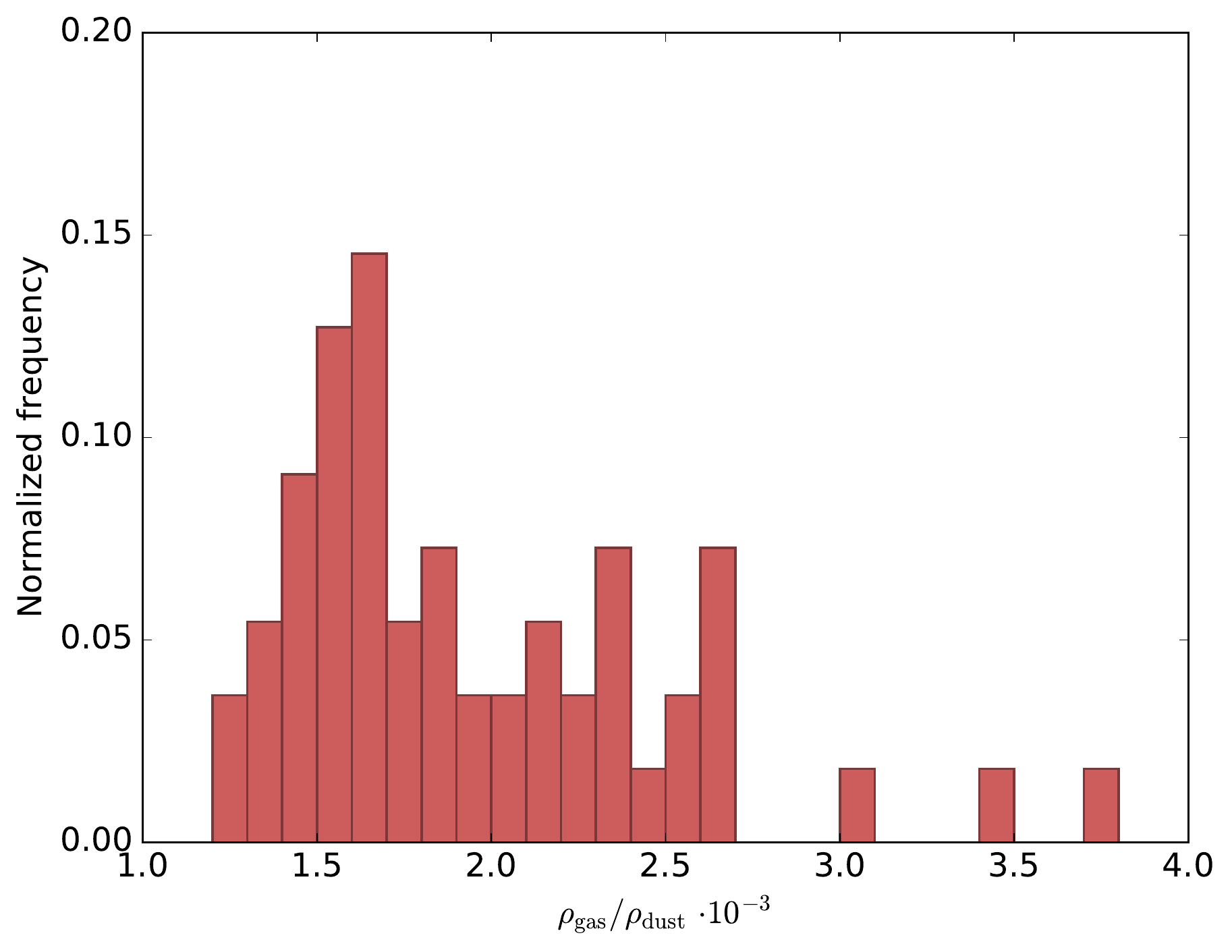} 
\caption{Degree of condensed Si (left panel) and gas-to-dust ratios by mass (right panel) for a subset of the DARWIN models of M-type AGB stars presented in \cite{bladh2015}.}
\label{f_dustprop}
\end{figure*}

Another important output from the time-series is grain properties. By noting the degree of material condensed into dust in the outer mass layers averaged over time we can calculate the average dust-to-gas ratio for each model. Figure~\ref{f_dustprop} shows the gas-to-dust mass ratios for a collection of oxygen-rich DARWIN models at solar metallicities.

\begin{figure*}
\centering
\includegraphics[width=\textwidth]{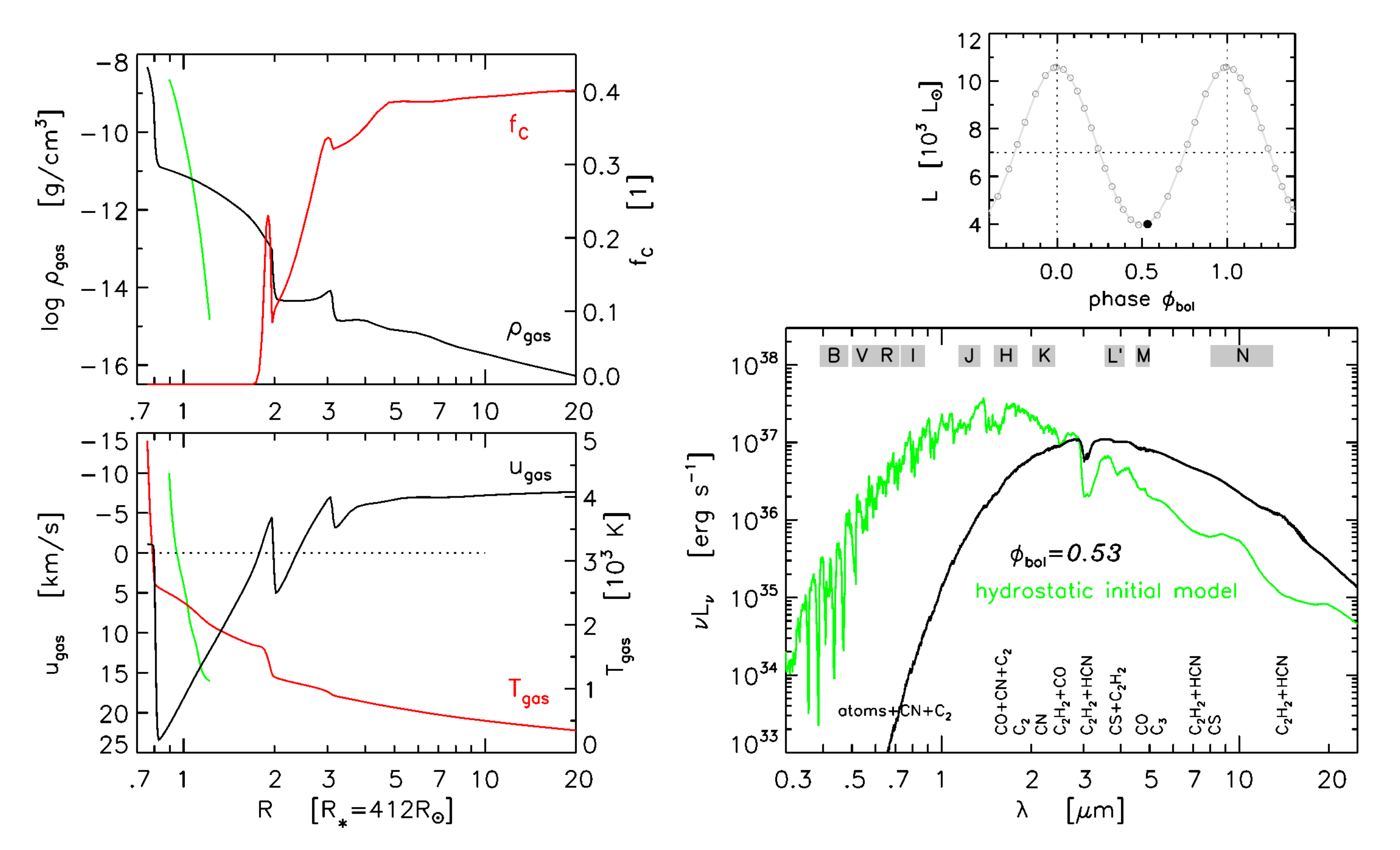} 
\caption{Radial profiles from an atmospheric snapshot of a DARWIN model of C-type (model S from \cite{nowotny11}), showing gas density and degree of condensed carbon in the top left panel, and gas velocity and temperature in the bottom left panel. The luminosity as a function of bolometric phase is shown in the top right panel, and the spectrum at a bolometric phase of 0.53 is shown in the lower right panel. The corresponding properties for the initial static model is shown in green. Image by courtesy of W. Nowotny.}
\label{f_nowotny}
\end{figure*}

The snapshots of the atmospheric structure can themselves be used to study physical processes in the extended atmosphere, i.e., maser emission or polarisation of light by dust. As an example of this, \cite{aronson17}
investigated the polarisation of light by dust at different phases and wavelengths using DARWIN models for M-type AGB stars. Processes affected by the shocks in the inner atmosphere, like non-equilibrium chemistry, can potentially also be studied using DARWIN models.

Detailed \textit{a posteriori} radiative transfer calculations of the atmospheric snapshots, using opacity sampling program COMA \citep{aringer2016}, can produce synthetic spectra, photometry and interferometric visibilities that can be compared with observations directly. Figure~\ref{f_nowotny} shows an example of a snapshot of the atmospheric structure and the corresponding spectrum from a C-type DARWIN model. Comparisons between synthetic and observed photometry have been made for DARWIN models of both C-type and M-type \citep{nowotny11,eriksson14, bladh2015}. Individual CO rotation-vibration lines, that depend on the shock dynamics, can be synthesised and compared to observations to probe the structure of the inner atmosphere. Such studies have been done for both M-type and C-type AGB stars \citep{nowotny11,Liljegren16,Liljegren17}. A first attempt at comparing high-resolution spectra from DARWIN models of M-type and the X-shooter Spectral Library was presented in the contributed talk ``O-rich LPVs in the X-shooter Spectral Library'' by Lancon at IAU Symp. 343. Examples of comparisons between observed and synthetic interferometric visibilities can be found in \cite{sacuto13,bladh17} for M-type AGB stars, and in \cite{sacuto11,rau2017,witt2017} for C-type AGB stars. 

\section{Future plans}
There are two kinds of challenges lying ahead: one is to further develop the description in the DARWIN models and the other is to increase the parameter space of the current model grids and producing model grids at different metallicities. 

The plan for the immediate future concerning model development is to make a more complete model for M-type AGB stars, by constructing a model version that includes the growth of both alumina and silicates, and allows for a variable Mg/Fe ratio in the silicate grains. With a more complete model for M-type AGB stars, the steps towards a model for S-type AGB stars become more feasible. A bit further down on the to-do list is the plan to include SiC dust in the DARWIN models for C-type AGB stars. There is also an ongoing effort to explore if it is possible to improve the DARWIN models to account for non-symmetrical effects \cite[see, e.g.,][and her talk ``Lumpy stars and bumpy winds'' at IAU Symp. 343]{liljegren18}.

\begin{figure*}
\centering
\includegraphics[width=\textwidth]{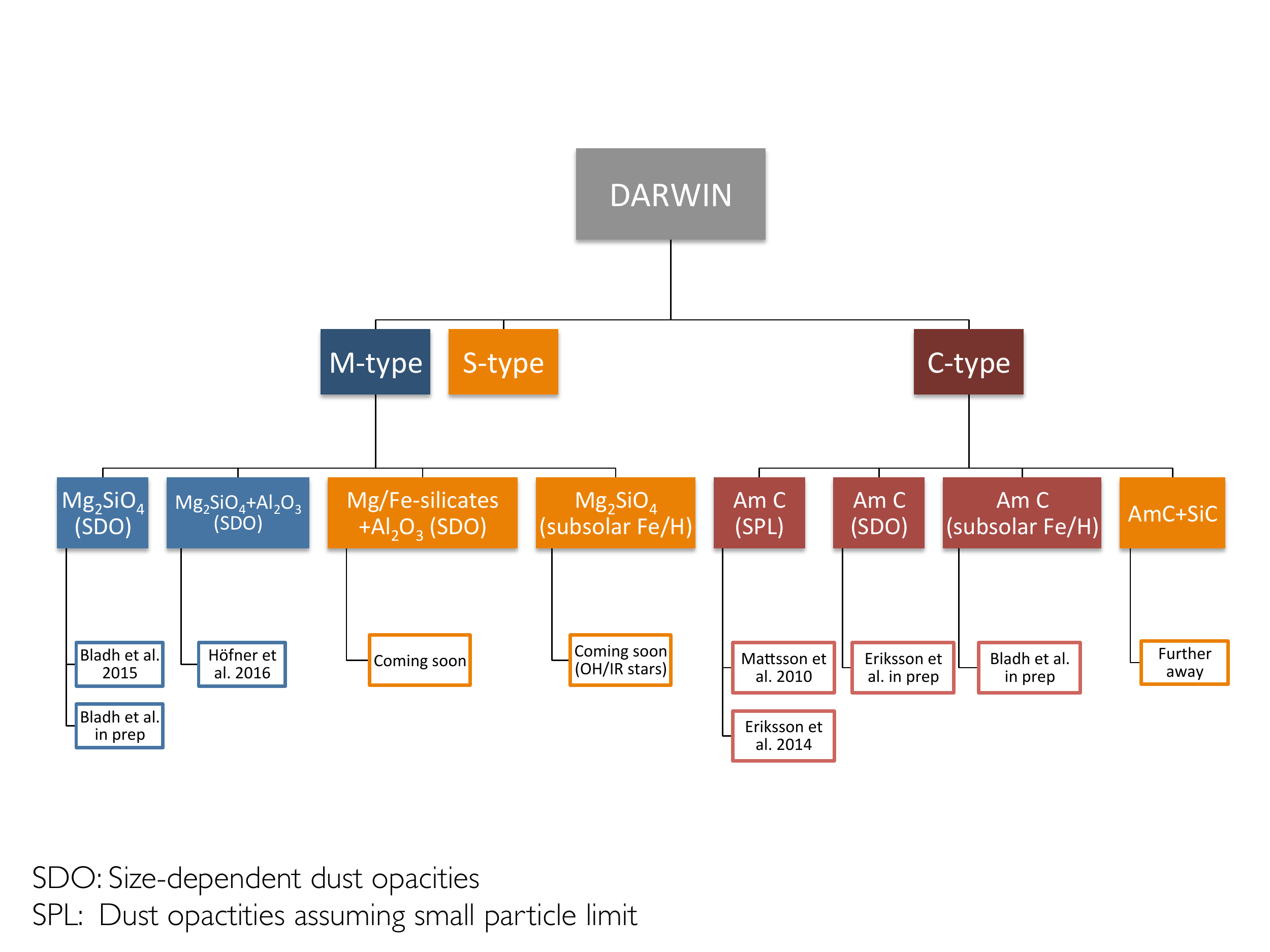} ´
\caption{Overview of DARWIN models currently available or to be published soon, with potential future plans marked in orange.}
\label{f_future}
\end{figure*}

Plans exist for expanding the model grids to include more extreme stellar parameters, but also to other metallicities. This is needed for a better understanding of stellar evolution during the AGB phase, as well as dust production in the early Universe. Figure~\ref{f_future} shows the potential future plans marked in orange, in addition to the DARWIN models currently available or soon to be published.

\begin{discussion}

\discuss{Decin}{The plots in which you show that the mass-loss rate and wind velocity are independent with respect to metallicity only for carbon-rich stars or for both carbon and oxygen-rich stars?}

\discuss{Bladh}{The result that the mass-loss rate is independent of metallicity for a sample of models at solar and sub-solar metallicity, if we compare models with the same input parameters and the same carbon excess, is for carbon stars only. Carbon stars manufacture the constituents needed for the wind-driving dust species internally (namely carbon), irrespectively of the metallicity environment they are in. The wind-driving dust species in M-type AGB stars, however, requires minerals, and a lower abundance of minerals will affect the ability to produce stellar winds at lower metallicities.}

\discuss{Sahai}{How do the wind parameters get affected as one goes towards an S-type composition, i.e. close to C/O=1? Since S-type stars are known to have strong winds, e.g., X Cygni.}

\discuss{Bladh}{There seems to be an ongoing debate of what C/O-ratios S-type AGB stars actually have and it may be as low as C/O=0.5 for some early-type AGB stars \citep{vaneck2018}. We have managed to produce winds for models at C/O=0.7, but this was not a systematic investigation. Theoretically, if S-type AGB stars all have C/O-ratio close to 1, then it will not be possible to drive a wind with the current DARWIN models.}

\discuss{Question}{You shared pictures of irregular dust grains (amorphous carbon, Mg$_2$SiO$_4$) with sharp edges. What approximation or method do you use for calculation of the dust optical properties - e.g. spherical grains, ellipsoidal grains or other?}

\discuss{Bladh}{The time-dependent description of grain growth in DARWIN models follows the method of \cite{gail1999}. We assume spherical grains, and calculate the dust opacity with Mie theory. Both dust-species are assumed to be amorphous (amC and Mg$_2$SiO$_4$). The optical data is taken from \cite{rolma91,jaeger2003}.}

\end{discussion}

\end{document}